\documentstyle [12pt] {article}
\begin{document}
\begin{center}
{\bf A Possible Theoretical Model For Studying Superconductivity In Fe-based Systems}\\
\end{center}
\vspace* {1 cm}
\begin{center}
{\bf Ranjan Chaudhury}\\
{\bf S.N. Bose National Centre For Basic Sciences, Salt Lake, Calcutta (Kolkata)- 700098}\\
{\bf India}\\
\end{center}
\vspace* {1 cm}
\section {Abstract}

  A theoretical approach with a microscopic model is proposed for the observed "high temperature superconductivity" in the Iron-based compounds. The above scheme takes into account two important aspects $viz.$ ($i$) superconducting transition close to magnetic ordering and ($ii$) the layered structure. From the calculation of the superconducting transition temperature, it is shown that in the $Fe$-based superconductors the magnetic mechanism for superconductivity, operating through the effective attractive Coulomb interaction within the framework of the Fermi Liquid theory, is highly plausible.\\
\vspace* {10 cm}
\section {Introduction}
  There has been a lot of excitement over the recent discovery of superconductivity in a new class of systems viz. the doped phase of layered rare earth-transition metal based compounds in the temperature regime of $25$-$50$K. They have a lot of similarities with the high temperature superconductors belonging to the Cuprates family in certain important experimental features. These include ($i$) layered structure with specific roles of various layers and ($ii$) the proximity to a magnetically ordered state. In this article we will only focus on the systems which contain $Fe$ as the transition metal component. \\

  These $Fe$-based superconductors can be divided into $3$ types depending on their chemical composition. They are :- ($1$) $RO_{1-x}F_{x}FeAs$ where $R$ represents a rare earth element like $La$, $Sm$ etc.; ($2$) $X_{1-x}Y_{x}Fe_{2}As_{2}$ where $X$ stands for any bivalent metal like $Ca$, $Ba$, $Sr$ etc. and $Y$ represents a monovalent metal like $Na$, $K$ etc. and ($3$) $XFe_{1-x}Y_{x}AsF$ where $X$ implies a bivalent metal and $Y$ is a transition metal different from $Fe$.  The experiments have shown that in all these systems the $Fe$-$As$ layer is responsible for both electrical transport as well as magnetism and that the magnetism is of itinerant nature [1,2]. In the parent or undoped phases, $i.e.$ the phases corresponding to $x=0$, all the above systems are insulator and exhibit long range magnetic ordering of spin density wave type. For the systems belonging to the subclass ($1$) above , the rare earth-oxygen layer acts as a charge reservoir and provides electrons to the $Fe$-$As$ layer to act as carriers when $O$ is partially replaced by $F$. This turns the system conducting and also suppresses the long range magnetic ordering. The system behaves as a normal paramagnetic metal with enhanced spin correlations and becomes superconducting at lower temperatures. The superconductors belonging to subclass ($2$) are hole doped and those belonging to subclass ($3$) can be either electron or hole doped. The common feature exhibited by all these $3$ subclasses of systems is however emergence of superconductivity close to magnetic ordering under doping. \\

  It is worthwhile to point out that the behaviour of the normal conducting phases of these $Fe$-based systems is very close to that of a usual metal, unlike those of the several members of the Cuprate family. Moreover, the complete phase diagram of these $Fe$-based systems as a function of doping, is much less complex compared to those of the Cuprates. Besides, from the phenomenology it seems that the roles of phonons and other electronic charge excitations on superconductivity is rather unimportant [3].\\

  On the basis of these above observations, we propose a simple microscopic model here for the superconducting phase of $Fe$-based superconductors. In our theoretical model we invoke a mechanism for electron pairing close to magnetic ordering, based on the dielectric function formalism [4]. Moreover, we also incorporate the crucial role played by the anisotropic layered structure in our calculations to some degree. We then explore the possibility of the effective Coulomb interaction between the electrons (or holes) becoming attractive due to the presence of the enhanced spin fluctuations in the system. Finally, we present an analytical formalism for calculating the superconducting transition temperature for our system, when the above mechanism is operative.\\

\section {Mathematical Formulation and Calculation of Superconducting Transition Temperature}
 
 We explore the pairing scheme of Bardeen-Cooper-Schrieffer (BCS) in a Fermi Liquid (FL) background. The assumption of the FL description of the normal phase is justified, as $r_s$ values of the equivalent electron (or hole) gas in $Fe-As$ layers in the doped phases of these systems fall in the range of $4$ to $5$, as determined from the experimental values of the carrier concentration and furthermore there is no experimental support for strong on-site Coulomb correlation either [5,6].\\

 ({\bf A}) We first propose a very generalized model which can be appropriate for these $Fe$-based superconductors. This is written as :-
\begin{equation}
 {\mathcal H}_{gen.} = {\sum_{m,{\bf k}}{\epsilon_{{\bf k},m}} c_{{\bf k}{\sigma},m}^{+}c_{{\bf k}{\sigma},m}} + {\sum_{{\bf k},{\bf k}^{\prime},m}V_{{\bf k}{\bf k}^{\prime},m}b_{{\bf k}^{\prime},m}^{+}b_{{\bf k},m}} +{\sum_{<mn>,{\bf k},{\bf k}^{\prime}}{\lambda}^{mn}_{{\bf k},{\bf k}^{\prime}}[b_{{\bf k}^{\prime},m}^{+}b_{{\bf k},n} + h.c.]}
\end{equation}
 where, in the right hand side the first and the second term together represent the standard BCS Hamiltonian corresponding to the intra-layer pairing with $m$ being the layer index for a particular $Fe$-$As$ layer. The ${\epsilon}({\bf k})$ dispersion is highly anisotropic with the major contribution coming from the intra-layer kinetic energy. The $c$'s are the usual single fermion operators and the $b$'s are the Cooper pair operators. The last term represents the hopping of Cooper pairs between two successive $Fe$-$As$ layers. It may be recalled that $b$'s are related to the $c$'s in the following way,
\begin{equation}
{b_{\bf k}} = c_{{-\bf k}{-\sigma}}c_{{\bf k}{\sigma}}
\end{equation}
  The basic idea is that the Cooper pairs are formed in each of the above mentioned layers through an effective attractive interaction $V_{{\bf k}{\bf k}^{\prime},m}$ ; however since superconductivity is a $3$-dimensional phenomenon, we had to invoke an inter-layer process coupling the pairs from different layers with a parameter $\lambda$. This inter-layer process may also contain a screened Coulomb interaction due to the background dielectric constant (with contributions from the layers containing the rare earth atoms), accompanying the pair tunnelling. It should be stressed that the parameter $\lambda$ is expected to be much smaller compared to $V$. Furthermore, we assume that the intra-layer attractive pairing interaction is a function of the in-plane momentum transfer of the electrons $i.e.$ ${{\bf k}_{||}-{\bf k}^{\prime}_{||}}$ only. The pairing interaction can be determined from the longitudinal dielectric function of the $2$d electron (or hole) gas, which can take into account the spin fluctuations.  Besides, the out of plane components of the electron momenta are taken to be box quantized in the following form \\
\begin{equation}
{\bf k}_{perp.} = {\frac{2{\pi}p}{L}}
\end{equation}
 where $p$ is an integer, +ve or -ve including zero and $L$ is the size of the lattice in the $z$ direction.\\ 

  We could define the order parameters $viz.$ the BCS gap function and the pair tunnelling function in the following way
\begin{equation} 
{\Delta}_{\bf k} = -{\sum_{{\bf k}^{\prime},m}V_{{\bf k}{\bf k}^{\prime},m}<b_{{\bf k}^{\prime},m}^{+}>}
\end{equation}
 and\\
\begin{equation}
{T_{\bf k}} = {\sum_{{\bf k}^{\prime},m,n} {\lambda}_{{\bf k}{\bf k}^{\prime}}^{m,n}<b_{{\bf k}^{\prime},m}^{+}>}
\end{equation}

 Carrying out a standard mean field treatment on our Hamiltonian, we generate the coupled self-consistent equations for the above order parameters. Our preliminary calculations with these bring out interesting results for the Bogoliubov quasi-particle excitation spectra and the superconducting gap equation. The quasi-particle spectra determined for our anisotropic layered system, exhibits significant and non-trivial departure from the conventional one known for the isotropic $3$d system [7]. In particular, the expression for the excitation energy gap turns out to be slightly different from the BCS gap parameter. Detailed calculations are in progress and the results will be reported later. \\

 ({\bf B}) We now work with a simpler version of the earlier model $i.e.$ the generalized model given by equation ($1$). Here we consider only the first two terms from equation ($1$) and remove the explicit layer indices. Therefore our simplified model is similar to the BCS model for an isotropic system; however the pairing interaction $V$ is once again assumed to be a function only of the in-plane momentum transfer and is explicitly calculated from the response functions corresponding to an interacting $2$d electron (or hole) gas. In particular, this dielectric function includes exchange-correlation corrections beyond Random Phase Approximation (RPA) and can be related to the spin susceptibility [4,8]. The repulsive bare Coulomb interaction $V_{0}({\bf q})$ is also taken corresponding to the $2$d electron gas and is given by \\
\begin{equation}
 {V_{0}({\bf q})} = {\frac{2{\pi}e^2}{{\epsilon}_Bq}}
\end{equation}
 where, ${\epsilon}_B$ is the background dielectric constant of the system, which has the main contributions from the layers containing the rare earth ions, besides Oxygen and Flourine ions for some of the $Fe$-based superconductors.\\

 Let us focus on the relevant set of equations now. From the expressions for the static and wave-vector dependent charge and spin responses for $2$d electron (hole) gas, we arrive at the following important equations [4,8]\\
\begin{equation}
 {\epsilon}^{-1}(q) = 1 - V_{0}(q){\pi}(q);\
 {\pi}^{-1}(q) = {\pi}_{0}^{-1}(q) + V_{0}(q)[1-G_s(q)]
\end{equation}
 where ${\epsilon}(q)$, ${\pi}(q)$ and ${\pi}_0(q)$ are the longitudinal dielectric function, the full polarizability function and the RPA polarizability function respectively. The function $G_s(q)$ is the spin symmetric local field correction. The expression for ${\pi}_0(q)$ is given by [9] \\
\begin{equation}
{\pi}_0(q) = {\frac{N(0)}{|q^{\prime}|}}[|q^{\prime}| + 2sgn(\nu){\theta}({\nu}^2-1)({\nu}^2-1)^{\frac{1}{2}}]
\end{equation}
 where $|q^{\prime}| = + {\frac{|q|}{2k_F}}$; ${\nu} = -{\frac{|q|}{2k_F}}$; $k_F$ is the Fermi wavevector and  $N(0)$ is the single fermionic (electron or hole) density of states for one kind of spin at the Fermi surface. Similarly, we have the following expression for the magnetic spin susceptibility \\
\begin{equation}
{\mu}_B^2{\chi}_s(q) = {\frac{{\pi}_0(q)}{1-V_0(q)G_a(q){\pi}_0(q)}}
\end{equation}
 where $G_a(q)$ is the spin anti-symmetric local field correction.\\

 From the above set of equations ($7$) - ($9$) we arrive at the following relation connecting the electrical and the magnetic response [4,8]
\begin{equation}
{\epsilon}^{-1}(q) = [{\mu}_B^2{\chi}_s^{-1}(q) - V_0(q)(G_s(q)-G_a(q))]{\pi}^{-1}(q)
\end{equation}
 The above equation ($10$) brings out an extremely important quantum effect. It  shows that for the $q$ values where the magnetic spin susceptibility is highly enhanced over the magnitude expected from the RPA or Pauli response so that the first quantity (in the bracket) on the right hand side becomes negative, then the dielectric function corresponding to those $q$ modes can itself become negative. It can be argued that the function ${\pi}^{-1}(q)$ is always positive because of the stability criterion [4]. This has the immediate consequence for the effective static Coulomb interaction given by $V_{eff}(q) = {\epsilon}^{-1}(q) V_0(q)$ $viz.$ the effective interaction becoming attractive corresponding to the $q$ modes where a paramagnetic layer exhibits enhanced magnetic spin susceptibility. This further leads to a scenario where the Cooper pairing can take place through this electronic mechanism itself. \\

 Let us now relook at the experimental situation for the $Fe$-based superconductors. The parent systems order magnetically with a long range SDW with $q$ of the order of $k_F$. In the doped phase which is paramagnetic and metallic, the magnetic spin susceptibility is highly enhanced in the above $q$ regime, particularly when the doping level is not too high. The results from neutron scattering experiments and ac susceptibility measurements confirm this [10]. Therefore, our proposed electronic mechanism for superconductivity in the $Fe$-based systems is indeed very very realistic.\\

 We would now like to explore the feasibility of such a mechanism to yield the superconducting transition temperature ($T_c$) in the range as observed for these superconductors.\\   

 In the weak coupling BCS theory for superconductivity in the $s$-wave channel, the attractive coupling constant is obtained by averaging the static $q$-dependent attractive interaction over the Fermi surface [4,11]. The equation for $T_c$ is related to this coupling constant and the cutoff energy (temperature) in the BCS square well model for the appropriate pairing interaction. In our case, this interaction is of the attractive Coulombic type and thus the cutoff scale is the Fermi energy (temperature).\\

 The attractive coupling constant (${\lambda}_e$) in our case is given by
\begin{equation}
{{\lambda}_e} = {\int_0^{{2k_F}_{||}}}{dq}{q} V_{eff}(q)
\end{equation}
 provided the right hand side of the above equation is negative; ${k_F}_{||}$ is the Fermi wavevector corresponding to the effective $2$d electron gas, which is related to the magnitude of the corresponding $r_s$, referred to earlier. \\

 The equation for $T_c$ under the above mentioned mechanism, then becomes
\begin{equation}
{T_c} = 1.13{T_{F}}{exp(\frac{-1}{|{\lambda}_e|})}
\end{equation}
 where $T_F$ is the Fermi temperature of the system.\\

 Assuming the value of $E_F$ to be between $0.6$ ev to $1.0$ ev for these $Fe$-based systems in the metallic phases, as determined from the photoemission experiments [6], we get from the above equation ($12$) the theoretical range of the magnitude of the attractive Coulomb coupling constant lying between $-0.1$ to $-0.2$ corresponding to the regime of observed $T_c$ in the broad spectrum of $25 K$ to $50 K$. Thus it is quite possible to attain the transition temperatures in this class of superconductors by invoking the magnetic mechanism in the weak coupling regime. \\

 We now examine more carefully the realisation of this strength of the attractive coupling from the microscopic many-body parameters used in our analysis earlier. Detailed calculation involving equations ($6$)- ($11$) and making use of the results from the earlier works [4,8,9], it can be shown that an average magnetic spin susceptibility enhancement over the $q$-space of the order of $10$ or even slightly less, can easily enable us to reach the above magnitude of the attractive Coulomb coupling constant. In this study, the typical bare repulsive Coulomb coupling constant (obtained at the RPA level, neglecting the local field corrections) was taken around +$0.3$ and the $q$-space averaged ratio of the very important and crucial correlation parameter $V_0(q)[G_s(q)-G_a(q)]$ and the RPA spin susceptibility, was kept within the allowed range for the metallic density limit. This theoretically obtained magnitude for the averaged spin susceptibility enhancement is very plausible, considering the fact that the strongest enhancement in these experimental systems takes place for the modes with $q$ around $k_F$, as the parent systems exhibit the long range SDW ordering. \\

\section{Conclusions}
 In this article, we have tried to present a simple but realistic microscopic model and approach for investigating the superconductivity in the new class of $Fe$-based superconductors.We have proposed a magnetic mechanism for this layered system and have considered the conventional $s$-wave pairing. An approach very similar to the present one was applied by us successsfully to a ferromagnetic superconductor $Y_{9}Co_{7}$ earlier [4]. The scheme presented here, can easily be extended for exploring the anisotropic pairing channels within the BCS or Eliashberg scheme as well.\\

 We have made use of the many-body parameters extracted from the electron gas calculations. More accurate quantitative prediction can only be made by taking into account the detailed band structure effects.\\

 Many of the member systems in this family exhibit a structural transition in the vicinity of the long range magnetic ordering in the undoped parental phase [1-3]. This gives rise to the possibility of the availability of soft phonons even in the doped metallic phases. These phonons may also play a secondary role in mediating  the pairing interaction, besides the major contribution coming from the magnetic mechanism described here.\\

 Our theoretical approach can be strengthened by more detailed experiments. In particular, a  study of the frequency integrated spectrum of the dynamical structure factor $S({\bf q}, {\omega})$ in the doped phases, extracted from the inelastic neutron scattering experiments can throw lot of valuable information regarding the $q$-space averaged spin susceptibility enhancement.\\

\section{References}

$[1]$ Kamihara Y et al 2008 J.Am. Chem. Soc. 130 3296; Norman M R 2008 Physics 1 21 \\
$[2]$ C de la Cruz et al 2008 Nature 453 899; Xiao Y et al 2008 Cond-mat arxiv: 0811.4418v2 \\
$[3]$ Valenzuela B and Bascones E 2008, presented at LT 25 (Amsterdam, Netherlands)\\
$[4]$ Chaudhury R and Jha S S 1984 Pramana 22 431; Uspenskii Y 1979 Sov. Phys. JETP 49 822\\
$[5]$ Gang Mu et al 2008 Cond-mat arxiv:0806.1668v4 ; Pines D 1999 "Elementary Excitations in Solids" Chapter 3 (Perseus Books) \\
$[6]$ Zhang Y et al 2009 Supercond. Sci. Technol. 22 015007 (to appear); Li H et al 2008 Cond-mat arxiv:0807.3153v2 \\
$[7]$ Ginzburg V L and Kirzhnits D A 1982 "High Temperature Superconductivity" (New York: Consultants Bureau)\\
$[8]$ Asgari R et al 2003 Phys. Rev. B 68 2351161; Takayanagi K and Lipparini E  Phys. Rev. B 1995 52 1738\\
$[9]$ Chaudhury R and Gangopadhyay D 1995 Mod. Phys. lett. 9 1657;  Ruvalds J 1987 Phys. Rev. B 35 8869\\
$[10]$ Qiu Y et al 2008 Phys Rev. B 78 052508; Xiao Y et al 2008 Cond-mat arxiv:0811.4418v2\\
$[11]$ Schrieffer J R 2000 "Theory of Superconductivity" (Frontiers in Physics)\\

\end{document}